\documentclass[twocolumn,aps,showpacs,showkeys,prd,superscriptaddress,byrevtex]{revtex4-1}
\usepackage[dvipdfmx]{graphicx}
\usepackage{color}

\begin{document}

\title{Domain boundaries in Luttinger-Tisza ordered dipole lattices}

\author{S. Ashhab}
\affiliation{Qatar Environment and Energy Research Institute (QEERI), Hamad Bin Khalifa University (HBKU), Qatar Foundation, Doha, Qatar}
\author{M. Carignano}
\affiliation{Qatar Environment and Energy Research Institute (QEERI), Hamad Bin Khalifa University (HBKU), Qatar Foundation, Doha, Qatar}
\author{M. E. Madjet}
\affiliation{Qatar Environment and Energy Research Institute (QEERI), Hamad Bin Khalifa University (HBKU), Qatar Foundation, Doha, Qatar}

\date{\today}

\begin{abstract}
Motivated by the recent interest in the possible ordering of the CH$_3$NH$_3$ dipoles in the material CH$_3$NH$_3$PbI$_3$, we investigate the properties of domain boundaries in a simple cubic lattice of dipoles. We perform numerical simulations in which we set the boundary conditions such that the dipoles at opposite sides of the simulated sample are ordered in different directions, hence simulating a domain boundary. We calculate the lowest energy configuration under this constraint. We find that if we consider only dipole-dipole interactions the dipole orientations tend to gradually transform between the two orientations at the two opposite ends of the sample. When we take into consideration the finite spatial size of the CH$_3$NH$_3$ molecules and go beyond the point dipole approximation, we find that the domain boundary becomes sharper. For the parameters of CH$_3$NH$_3$PbI$_3$, our results indicate that the optimal energy structure has a boundary region of a width on the order of a single unit cell.
\end{abstract}

\maketitle

\section{Introduction}
\label{Sec:Introduction}

Since the pioneering work of Kojima {\it et al.} \cite{Kojima:2009aa} demonstrating a functional solar cell using methyl-ammonium lead iodide (CH$_3$NH$_3$PbI$_3$, also abbreviated as MAPbI$_3$) as the key component, there has been an ever increasing effort to take advantage of hybrid organic-inorganic perovskites (HOIPs) for energy harvesting \cite{Green:2017aa}, photoluminescence \cite{Kanemitsu:2017aa} and lasing \cite{Adjokatse:2017aa} applications. Indeed, HOIPs have emerged as a new class of materials with a high potential to challenge existing technologies. In this scenario, there has been a very strong experimental effort to develop efficient and stable devices \cite{Yang:2017ab,Chen:2018aa}. In parallel, the theoretical research aiming to provide basic understanding of many of the properties of HOIPs continues and includes mainly electronic structure studies based on density functional theory \cite{Even:2013aa,Umari:2014aa,Hu:2017aa}, molecular dynamics simulations aiming to characterize the ionic degrees of freedom \cite{Carignano:2016aa,Carignano:2017aa,Mattoni:2017aa,Lahnsteiner:2018aa}, and other theoretical approaches \cite{Frost:2014ab,Even:2016aa,Ashhab:2017aa,Madjet:2017aa,Li:2018aa,Li:2018ab}.

Even though there are many variations on the chemical formulations, MAPbI$_3$ is still the prototypical HOIP material. The high temperature phase, which is the stable phase above 327 K \cite{KNOP:1990aa}, shows a cubic arrangement of the inorganic cage and a dynamic behavior of the molecular cations that exhibit rotational degrees of freedom characterized by a relaxation time of a few picoseconds \cite{Carignano:2015aa,Leguy:2015aa,Bakulin:2015aa}. The tetragonal phase also shows some rotational degrees of freedom, but other phenomena have been observed that suggest the possibility of the formation of rotational glasses \cite{Fabini:2016aa}. The low temperature phase is orthorhombic, with all the cation orientations locked to an ordered arrangement dictated by the inorganic cage. Besides the kinetics of the cations, the inorganic cage is subject to interesting dynamics. In particular, the crystallographic characterization of the cubic phase shows that each iodine atom occupies a highly anisotropic ellipsoid with its minor axis along the line connecting the two Pb atoms to which the iodine atom is bonded \cite{Baikie:2015aa}. The interpretation of this phenomenon is that the inorganic lattice undergoes deformations that can be described using an extended cell of $2 \times 2 \times 2$ stoichiometric units. These deformations are coupled to the rotation of the organic cations \cite{Even:2016aa}. All the degrees of freedom of the system can be accounted for by performing molecular dynamics simulations. However, the complexity of the HOIPs is such that in order to have a complete representation of the problem it is necessary to have large supercells and long trajectories obtained with a reliable model. We have performed in previous work extensive first-principles simulations of MAPbI$_3$ on $4 \times 4 \times 4$ supercells to asses the structural properties of the ground state system \cite{Carignano:2017aa}. Excited state properties, such as hot carrier dynamics and its relaxation through electron-phonon coupling, were studied using smaller model systems \cite{Madjet:2017aa,Li:2018ab}. For the former, the results display structural (i.e. pair distribution functions) and kinetic properties (i.e. cation rotation timescale) in line with the experimental evidence, but also show the shortcomings of the simulation methodology. At high temperatures the coupling between molecular orientations and lattice deformations are expected to be weaker than at lower temperatures. At 450 K, we observed a slight anharmonicity in the halide dynamics and a fast rotation of all MA molecules. When the temperature is reduced to 370 K, where the system should remain in the cubic phase, the  $4 \times 4 \times 4$ supercells transform spontaneously to an antiparallel alignment of the dipoles that is likely the result of the small supercell used in the simulations. A question that arises is how to properly characterize the importance of dipolar interactions in the cubic phase in the absence of a coupling to the inorganic lattice.

Each of the many interesting properties of MAPbI$_3$ can be explained by one or some of the internal degrees of freedom in the material. For example, hysteresis in the electrical response of perovskite devices has been linked to ion migration\cite{Elumalai:2016aa} and also to formation and release of interfacial charges at the cell electrodes \cite{Weber:2018aa}. However, recent detailed experimental work strongly suggests that the material is a true ferroelectric \cite{Rakita:2017aa,Garten:2019aa}. Consequently the emergence of different orientational domains is expected to be  a dynamical one affected by electrical cycling \cite{Frost:2014aa}. The structure of the domain boundaries has also been linked to the high charge conductivity \cite{Rashkeev:2015aa,Rossi:2018aa}. Therefore, the understanding of the principles governing the structure of domain boundaries is important and related to the operation of devices.

In this paper, we focus on the role of the dipolar interactions between the molecular cations in determining the structure of the interfaces between regions with different dipolar arrangements. For this purpose, we employ a minimal model that describes the interaction between the dipoles in the lattice. In other words, we focus on the MA molecular dipoles and assume that their locations are fixed in space, hence ignoring the inorganic lattice formed by the Pb and I atoms. Although an accurate model of the material would have to include the inorganic lattice, the simplified model allows us to obtain clear insight about the role of the interactions between dipoles in determining the dipolar order and properties of dipolar domain boundaries. As we shall show below, the structure of the inter-domain region depends on the extended nature of the charge distribution of the MA cations. We include this finite extension of the dipoles in our calculations.

\section{Energy and optimal configuration}
\label{Sec:GroundState}

We consider a model where point dipoles (i.e.~dipoles of infinitesimal spatial size) are arranged in a simple cubic crystal structure. Each dipole is free to point in any direction. The dipoles interact with each other via the dipole-dipole interaction, which gives to the total energy
\begin{equation}
E = \frac{E_0}{2} \sum_{i,j} \frac{\vec{p}_i\cdot\vec{p}_j - 3 \left(\vec{p}_i\cdot\hat{r}_{ij}\right) \left(\vec{p}_j\cdot\hat{r}_{ij}\right)}{\left|r_{ij}\right|^3},
\label{Eq:DipoleDipoleInteractionEnergy}
\end{equation}
where $E_0$ is the characteristic dipole-dipole interaction energy scale given by
\begin{equation}
E_0 = \frac{d^2}{4\pi\epsilon_0 r_0^3},
\end{equation}
$\vec{p}_i$ is the unit vector in the direction of dipole $i$, $\hat{r}_{ij}$ is the unit vector pointing from the location of dipole $i$ to the location of dipole $j$, $r_{ij}$ is the distance between dipoles $i$ and $j$ measured in units of the lattice parameter, $d$ is the magnitude of the dipole moment, $\epsilon_0$ is the permittivity, and $r_0$ is the lattice parameter. The summation indices $i$ and $j$ run over all the dipoles in the material, excluding the terms with $i=j$. The factor of 2 in Eq.~(\ref{Eq:DipoleDipoleInteractionEnergy}) is used to avoid double-counting dipole pairs. We note here that our derivations and analysis below can be straightforwardly adjusted to describe other types of interactions, such as the Heisenberg interaction
\begin{equation}
E_{\rm Heisenberg} = \frac{E_0}{2} \sum_{\langle i,j \rangle} \vec{p}_i\cdot\vec{p}_j,
\label{Eq:HeisenbergInteractionEnergy}
\end{equation}
where the brackets in ``$\langle i,j \rangle$'' indicate that only nearest-neighbor pairs are included in the sum. However, since the dipole-dipole interaction model is expected to give a more accurate description of interactions in HOIP, we focus on this model here.

The statistical problem of dipole lattices has been theoretically addressed many years ago in the context of studies on the dielectric constants for crystals \cite{Lorentz}, and several summation strategies have been developed in order to obtain exact results \cite{Berlin:1952aa,Lax:1952aa,Nijboer:1958aa} in spite of the long-range nature of dipole-dipole interactions. Our interest here is focused on the properties of the optimal configuration (i.e.~the system configuration that minimizes its total energy), which should be qualitatively independent of issues related to the convergence of infinite sums with increasing summation range, and we therefore use a simple approach based on a spherical cutoff of the interactions. The optimal configuration of a homogeneous, single-domain system has been investigated in detail in the literature \cite{Luttinger:1946aa,*Luttinger:1947aa,Johnston:2016aa}.

The most energetically favorable combination of orientations for a pair of dipoles is to have the dipoles pointing in the same direction parallel to the line connecting the dipole locations. Obviously this situation cannot be realized for all pairs in a lattice of more than one dimension, because each dipole interacts with a large number of other dipoles that are distributed in different directions around it. A less optimal, but still favorable, configuration for a dipole pair is that in which the dipoles point in opposite directions perpendicular to the line connecting the two dipoles. The above two rules for favorable dipole alignment suggest a configuration in which one plane (for example, the xy plane) has an antiferromagnetically ordered dipole configuration with the dipoles all pointing perpendicular to the plane and the same pattern is repeated for all other parallel planes in the material. 

The above-described state is six-fold degenerate because of the equivalence of the three main axes of the lattice and the inversion symmetry. It was shown by Luttinger and Tisza that there are in fact an infinite number of energetically equivalent optimal configurations \cite{Luttinger:1946aa}. Any specific one of these configurations can be described as follows: the dipole at the origin points in some direction on the unit sphere with Cartesian components $p_x$, $p_y$ and $p_z$, with all directions being allowed. The orientations of all other dipoles are then determined according to a simple rule: the dipole at location $(x, y, z)$ [these being three integers] has Cartesian components $(-1)^{y+z} p_x$, $(-1)^{x+z} p_y$ and $(-1)^{x+y} p_z$. This structure has periodicity 2 in the x, y and z directions, and the periodicity in any of these three directions becomes 1 if the dipoles are oriented along that direction. The infinite degeneracy in the optimal configurations means that we can choose out of an infinite number of boundary conditions when we analyze the properties of domain boundaries between dipolar domains, because any of the energetically equivalent optimal configurations can be used as the bulk state on one or the other side of the domain boundary.

It should be noted that the infinite-fold degeneracy is a consequence of the perfect cubic symmetry and vanishing spatial size of the dipoles. If for example we take a tetragonal or orthorhombic lattice in which the dipole spacing in one direction is smaller than those in the other two directions, the symmetry is broken. Neighboring dipoles located along the short-distance direction will have the strongest interactions, and they will have the priority to minimize their interaction energy by aligning themselves along that direction. All other dipole configurations will have a higher total energy, hence resulting in two-fold degeneracy for the optimal configuration. Similarly, if the dipoles have a finite spatial size, i.e.~a finite distance separating the positive and negative charges, there will be an energetic advantage for configurations in which the positive charge from one dipole and the negative charge from a neighboring dipole minimize their energy by coming as close as possible to each other. This symmetry-breaking contribution to the energy appears at the energy scale $E_0\times\left(r_d/r_0\right)^2$, where $r_d$ is the intra-dipole distance. With this contribution to the energy taken into account, and assuming cubic symmetry of the lattice, the system has the six-fold degeneracy described above, where the optimal configuration has the dipoles oriented along one of the cubic crystal axes x, y or z.

\section{Domain boundaries}
\label{Sec:DomainBoundaries}

Based on recent experimental results, it is believed that the dipolar order in MAPbI$_3$ varies in space and changes in time 
\cite{Vorpahl:2018aa,Carignano:2017aa}. It can therefore be expected that a sample of the material will contain a large number of domains and domain boundaries at any point in time. This property is the basis for our motivation to study the properties of the boundaries between ordered domains that have different orientations.

\begin{figure}[h]
\includegraphics[width=7.0cm]{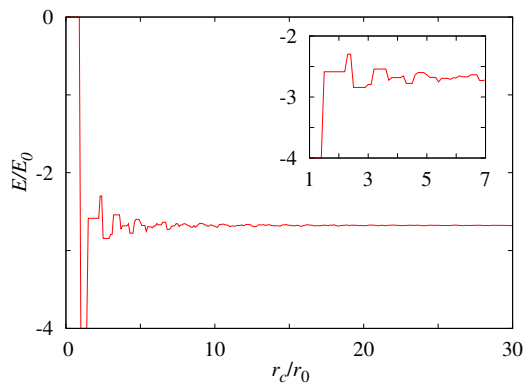}
\caption{Interaction energy per dipole (in units of the characteristic energy scale $E_0$) as a function of the interaction range $r_c$ (i.e.~the distance beyond which we ignore dipole-dipole interactions) measured in units of the lattice parameter $r_0$. For large values of $r_c/r_0$, the energy per dipole clearly converges to a value around $-2.7 E_0$ after the initial oscillatory behavior.}
\label{Fig:EnergyAsFunctionOfInteractionRange}
\end{figure}

We have performed numerical calculations based on Eq.~(\ref{Eq:DipoleDipoleInteractionEnergy}) to investigate these boundaries. We shall generally use an interaction range (or cutoff) $r_c$ of distance $3r_0$. With this value of the cutoff, the energy per dipole in a perfectly ordered material is $-2.79 E_0$. As we show in the Appendix, this value of the energy is only weakly dependent on the exact value of the cutoff, and taking a cutoff of $r_c/r_0\sim 30$ would give an optimal configuration energy close to $-2.68 E_0$ per dipole. When we simulate systems with a domain boundary, the interaction range sets a lower bound on the number of dipoles that we must include in the simulation, which is the reason why we use the relatively small value $r_c/r_0=3$. Importantly, however, our main results are essentially independent of the exact value of the cutoff. For example, although taking $r_c/r_0=3$ will produce energies that can deviate from the asymptotic values by a few percent (as can be seen in Fig.~\ref{Fig:EnergyAsFunctionOfInteractionRange}), this value of the cutoff is expected to give the correct boundary profile and functional dependence of the boundary energy.

To calculate the profile and energy of a domain boundary, we construct two $10 \times 2 \times 2$ dipole lattices with perfect order in each lattice. We attach the two lattices together to construct a $20 \times 2 \times 2$ lattice with a domain boundary in the middle. The small numbers of dipoles in the y and z directions are justified by the fact that translation symmetry is preserved perpendicular to the domain boundary and the lowest-energy dipole configuration will still have periodicity 2 in these directions. After initializing the dipole configuration as described above, we allow the dipoles to rotate and lower their energy. In principle, any minimization method can be used in this calculation to find the lowest energy in the space of the 160 variables that specify the dipolar configuration of the 80 dipoles in the sample. In our calculations we take any given not-yet-optimized configuration, calculate the torques felt by the dipoles as a result of their dipole-dipole interactions and slightly rotate each dipole in accordance with the direction and magnitude of the torque that it feels. We start with a relatively large step size for the updates from one configuration to the next, and whenever the energy stops decreasing we reduce the step size, until we reach an angular step size of $\sim 10^{-6}$. As mentioned above, in this calculation we do not include the inorganic lattice, and we therefore do not perform a density-functional-theory relaxation of the atomic positions in the lattice. We use periodic boundary conditions in all directions while fixing the orientations in the three farthest layers on each side of the boundary, i.e.~the three leftmost and three rightmost layers in the $20 \times 2 \times 2$ lattice (which is a total of 24 fixed dipoles). In this way, there is only one boundary where the dipoles will feel the effect of dipoles from the other domain and adjust their orientations to lower the energy.

\begin{figure}[h]
\includegraphics[width=8.0cm]{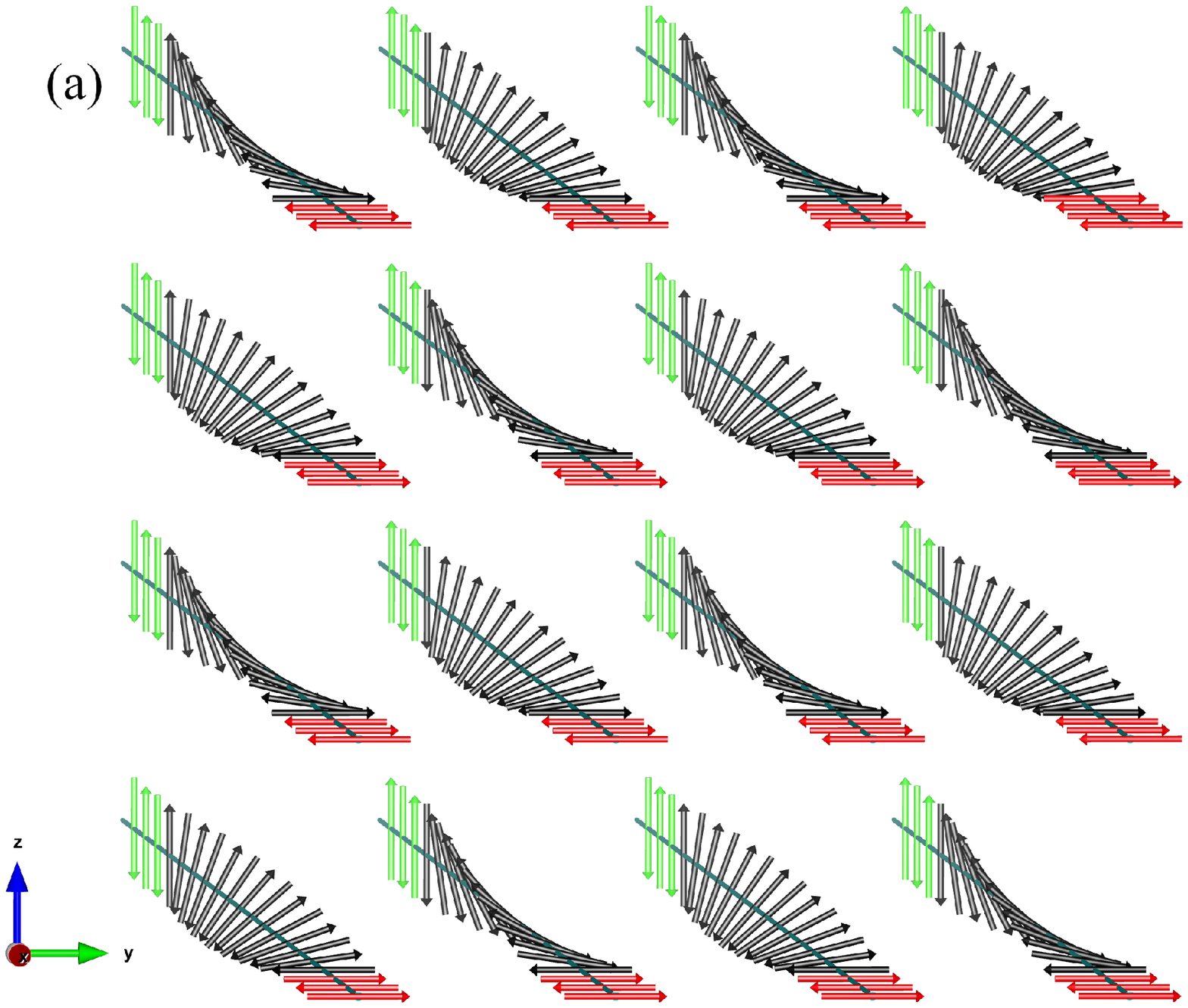}
\includegraphics[width=8.0cm]{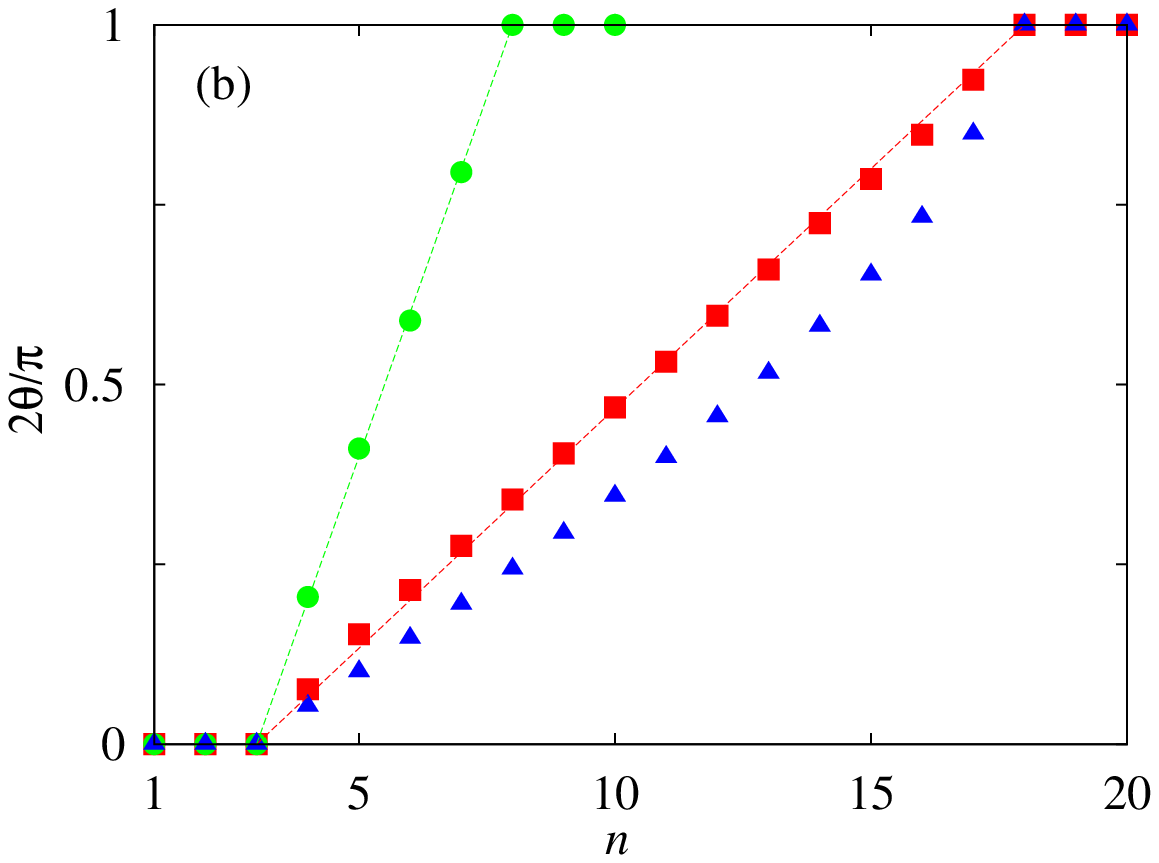}
\caption{(a) Dipole configuration of a boundary between two domains having dipole orientations that are perpendicular to each other and are both parallel to the plane of the domain boundary. The point dipoles are arranged in a simple cubic lattice; the lines that go through the arrows are added to indicate that the positions of the dipoles are exactly at the intersection points of these lines and the arrows that represent the dipoles orientations. In particular, we take the axis perpendicular to the domain boundary to be the x axis, the dipoles in the left domain to be aligned with the y axis (in the Luttinger-Tisza configuration described in the text) and the dipoles in the right domain to be aligned with the z axis. The simulation has 20 layers in the x direction, with three fixed layers on the far left, three fixed layers on the far right and 14 layers whose dipole orientations are free to rotate in order to find the minimum-energy configuration, which we show here. After optimization, the dipoles rotate gradually from one orientation to the other. Panel (b) shows the angle that the dipoles make with the y axis as a function of layer index. The red squares are taken from the configuration shown in Panel (a). The green circles show the results for a ten-layer simulation, i.e.~with four free layers. The blue rectangles show the results for the case when the dipoles in the right domain are aligned with the x axis instead of the z axis. The dashed lines are straight lines connecting the end points of the boundary. They show that in the case when the dipoles on both sides are parallel to the plane of the boundary, the angle increases almost linearly from one domain to the other. When the dipole orientations have a component perpendicular to the boundary, the change in angle is no longer linear.}
\label{Fig:BoundaryDipoleProfile}
\end{figure}

In Fig.~\ref{Fig:BoundaryDipoleProfile}(a) we show the results for the case where in one domain the dipoles are aligned with the y axis while in the other domain they are aligned with the z axis. In other words, both orientations are parallel to the plane of the domain boundary. We can see that the boundary is quite smooth: the orientation angle changing almost linearly as one moves from the fixed layers on one side to the fixed layers on the other side of the boundary, as shown more clearly in Fig.~\ref{Fig:BoundaryDipoleProfile}(b). The energy of the domain boundary is calculated by taking the difference between the total energy of the simulated sample containing a domain boundary and the total energy of a single-domain Luttinger-Tisza configuration:
\begin{eqnarray}
E_{\rm Boundary} & = & E_{\rm {Relaxed\ configuration\ with\ fixed\ boundaries}} \nonumber \\ & & - E_{\rm {Luttinger-Tisza\ configuration}}.
\end{eqnarray}
As the domain boundary energy is proportional to the cross-sectional area of the boundary, we shall plot the energy per unit area $E_{\rm Boundary}/A$, where we define $A$ as the number of unit cells (or dipoles) in the cross-sectional area of the simulated sample. In order to eliminate the second, artificial domain boundary that appears on the left and right boundaries of the simulated sample (i.e.~the boundary that appears because of the periodic boundary conditions used in our calculations), for the energy calculation we pad the simulated sample with a few layers that have the appropriate dipole orientations on both sides of the sample. For the configuration shown in Fig.~\ref{Fig:BoundaryDipoleProfile}, the energy of the domain boundary per unit cell area (i.e.~an area of $r_0^2$) is $0.018 E_0$. The fact that this energy is positive is a reflection of the fact that the unconstrained optimal configuration is one in which the whole sample contains a single domain. Any other configuration will have a higher energy, and the excess energy is identified as the energy of the boundary. Nevertheless, this boundary energy is remarkably small. To demonstrate how small this energy scale is, we note that the energy associated with flipping the direction of a single dipole in the optimal configuration is $4 \times 2.79 E_0 \sim 10 E_0$. The reason for this smallness is that in a smooth boundary the dipoles in any given region only slightly deviate from an energetically optimal configuration when seen locally.

\begin{figure}[h]
\includegraphics[width=8.0cm]{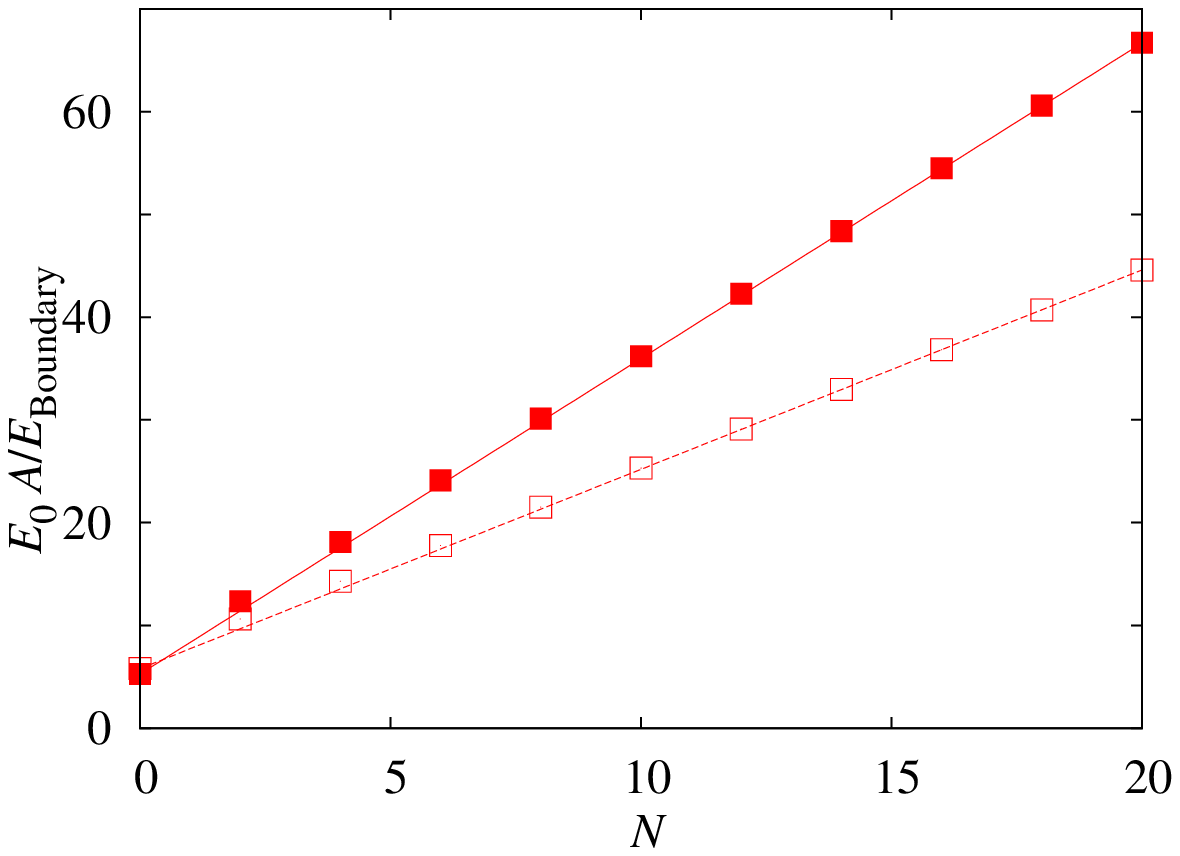}
\caption{Inverse of the domain boundary energy per unit cell area ($E_0 A/E_{\rm Boundary}$) as a function of the width of the boundary. The width $N$ is defined as the number of layers whose dipoles are allowed to rotate while the outer layers are fixed. The dipole orientations at the boundaries of the simulated sample are the same as in Fig.~\ref{Fig:BoundaryDipoleProfile}. The filled squares are obtained with an interaction cutoff distance of $r_c/r_0=3$, while the open squares are obtained with $r_c/r_0=10$. The lines are straight lines between the first and last points, and they show that the energy is inversely proportional to the width of the boundary, up to a constant. The extreme closeness of the two lines at $N=0$ is somewhat coincidental, as other values of the cutoff would give different y-intercepts, mostly in the range 5-15.}
\label{Fig:EnergyAsFunctionOfBoundaryWidth}
\end{figure}

\begin{figure}[h]
\includegraphics[width=8.0cm]{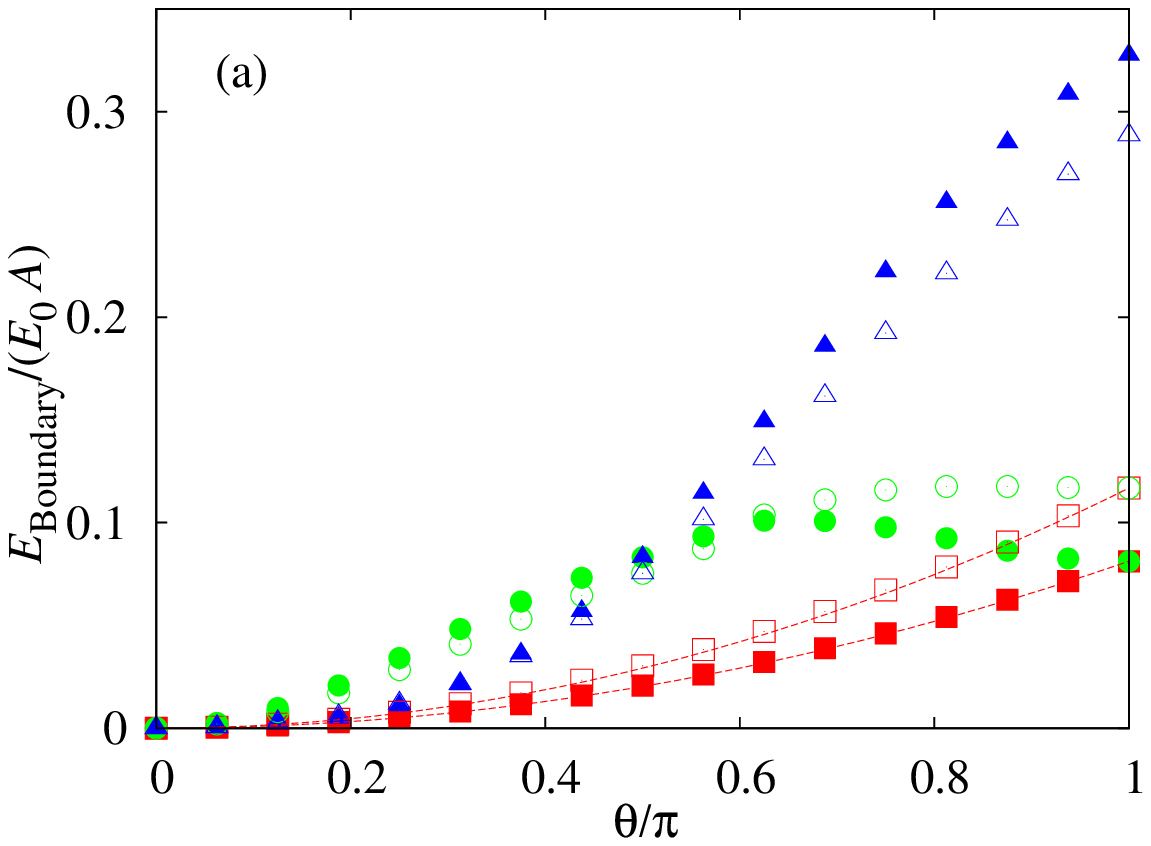}
\includegraphics[width=6.0cm]{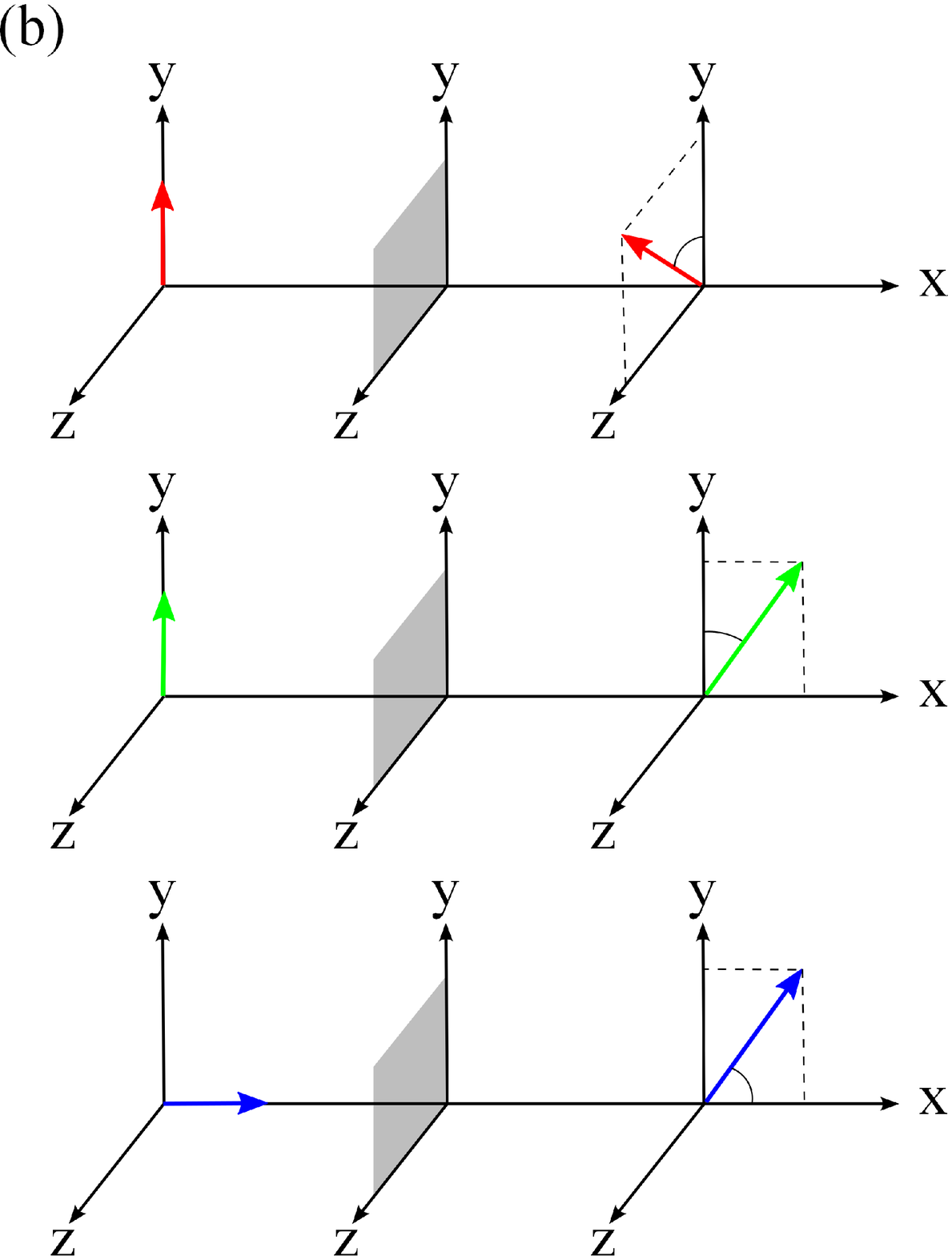}
\caption{(a) Energy of the domain boundary per unit cell area $E_{\rm Boundary}/(E_0 A)$ as a function of the mismatch angle between the orientations on the left and right sides of the boundary. The filled and open symbols correspond to interaction cutoffs $r_c/r_0$ of 3 and 10, respectively. Although in some cases the different cutoff values result in different energies, we have verified that the variation of the dipole orientations across the domain boundary is essentially independent of the cutoff. Panel (b) illustrates the three cases of orientation combinations that we analyze here. The boundary is always taken to be parallel to the yz plane. The three different cases are defined by the orientation in the left domain and the plane containing the orientation in the right domain: y and yz (red squares), y and xy (green circles), and x and xy (blue triangles). The dashed lines in (a) are quadratic functions between the first and last red data points, and they show that in this case the energy is proportional to the square of the mismatch angle $\theta$.}
\label{Fig:EnergyAsFunctionOfMismatchAngle}
\end{figure}

As can be seen in Fig.~\ref{Fig:EnergyAsFunctionOfBoundaryWidth}, the energy of the boundary is inversely proportional to the width of the boundary, up to a constant that can be ignored for large widths. This result is reasonable, because the rate of change in orientation angles as one goes through the boundary layers is inversely proportional to the number of layers in the boundary. For small orientation angle rotation rates, the energy of a given dipole is proportional to the square of the mismatch angle with neighboring dipole, i.e.~inversely proportional to the square of the width of the boundary. Given that the number of dipoles in the boundary region is proportional to the width, we find that the total energy of the boundary is inversely proportional to the width.

Next we perform more systematic calculations on domain boundaries for various combinations of dipole orientations. First we take the dipoles in one domain to be aligned with the y axis while the dipoles in the other domain lie in the yz plane and make an angle $\theta$ with those in the first domain. In other words, both orientations are parallel to the plane of the boundary. Note that $\theta$ ranges from 0 to $\pi$: when $\theta=0$ we have only one domain and hence no domain boundary, when $\theta=\pi/2$ we have the case shown in Fig.~\ref{Fig:BoundaryDipoleProfile} and discussed above, and when $\theta=\pi$ dipole pairs on opposite sides of the boundary (before allowing the dipoles to relax) are parallel instead of being anti-parallel to each other (the latter being the most energetically favorable situation). As can be seen in Fig.~\ref{Fig:EnergyAsFunctionOfMismatchAngle}, and as might be expected, the energy increases as the mismatch angle $\theta$ increases from 0 to $\pi$. The energy very closely follows a quadratic dependence on $\theta$. Other combinations of orientation directions give different results. When one of the domains is taken to have dipoles oriented perpendicular to the plane of the boundary (i.e.~the dipoles being oriented along x) and the other domain orientation lies in the xy plane, the functional dependence of the energy on $\theta$ is clearly no longer quadratic. Furthermore, the energy of the boundary is significantly higher than when the dipole orientations are all parallel to the plane of the boundary. This result indicates that bending the dipole orientations outside the plane of the boundary is associated with a higher energy than bending the orientation in the plane of the boundary. If we take the dipole orientation in the left domain to point in the y direction (i.e.~parallel to the plane of the boundary) and the dipole orientation in the right domain to be in the xy plane, the energy exhibits non-monotonic behavior, with a maximum at or close to $\theta=2\pi/3$. The non-monotonic dependence on the mismatch angle is also a result of the fact that the energetic cost of bending the dipoles depends on the direction of bending: a boundary with bulk orientations given by y and $-$y can have a lower energy than a boundary with bulk orientations y and x (in spite of having a larger mismatch angle) because the former case allows the dipoles to gradually change from y to z to y, hence avoiding to develop dipole components perpendicular to the plane of the boundary.

\begin{figure}[h]
\includegraphics[width=8.0cm]{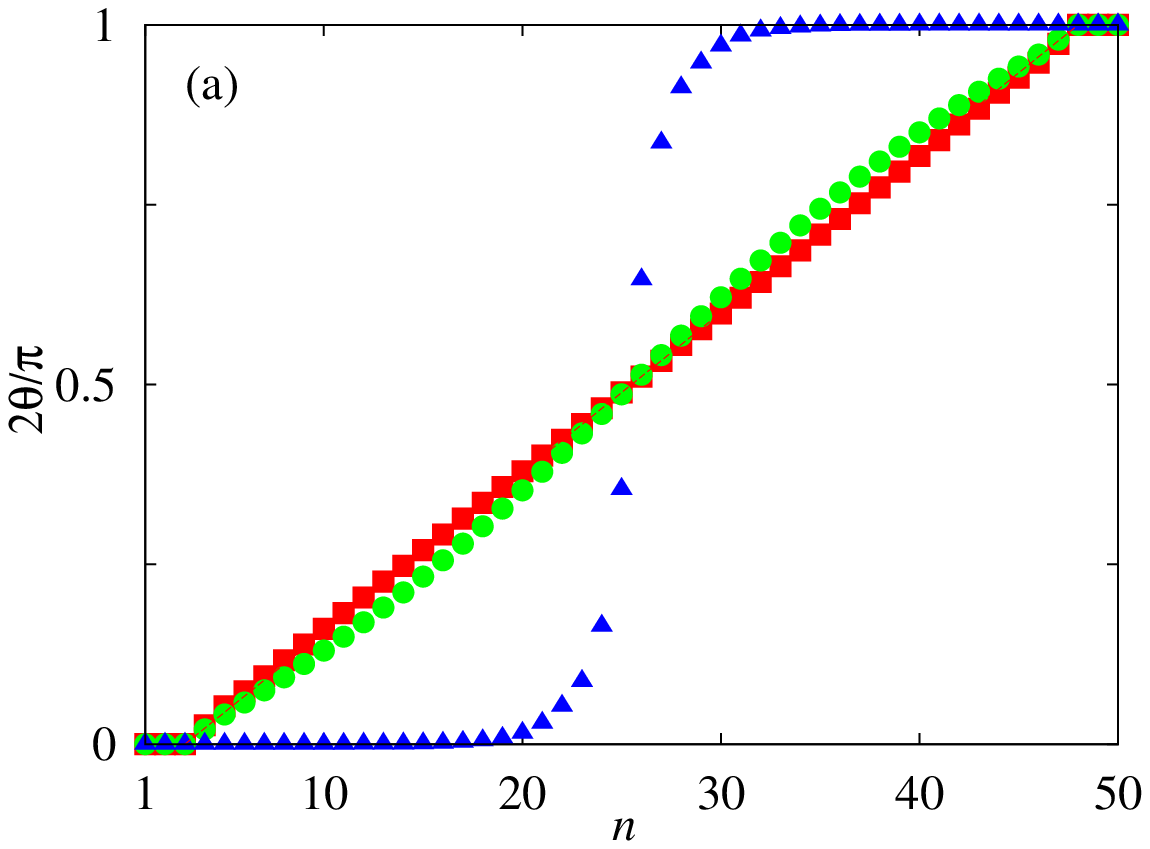}
\includegraphics[width=8.0cm]{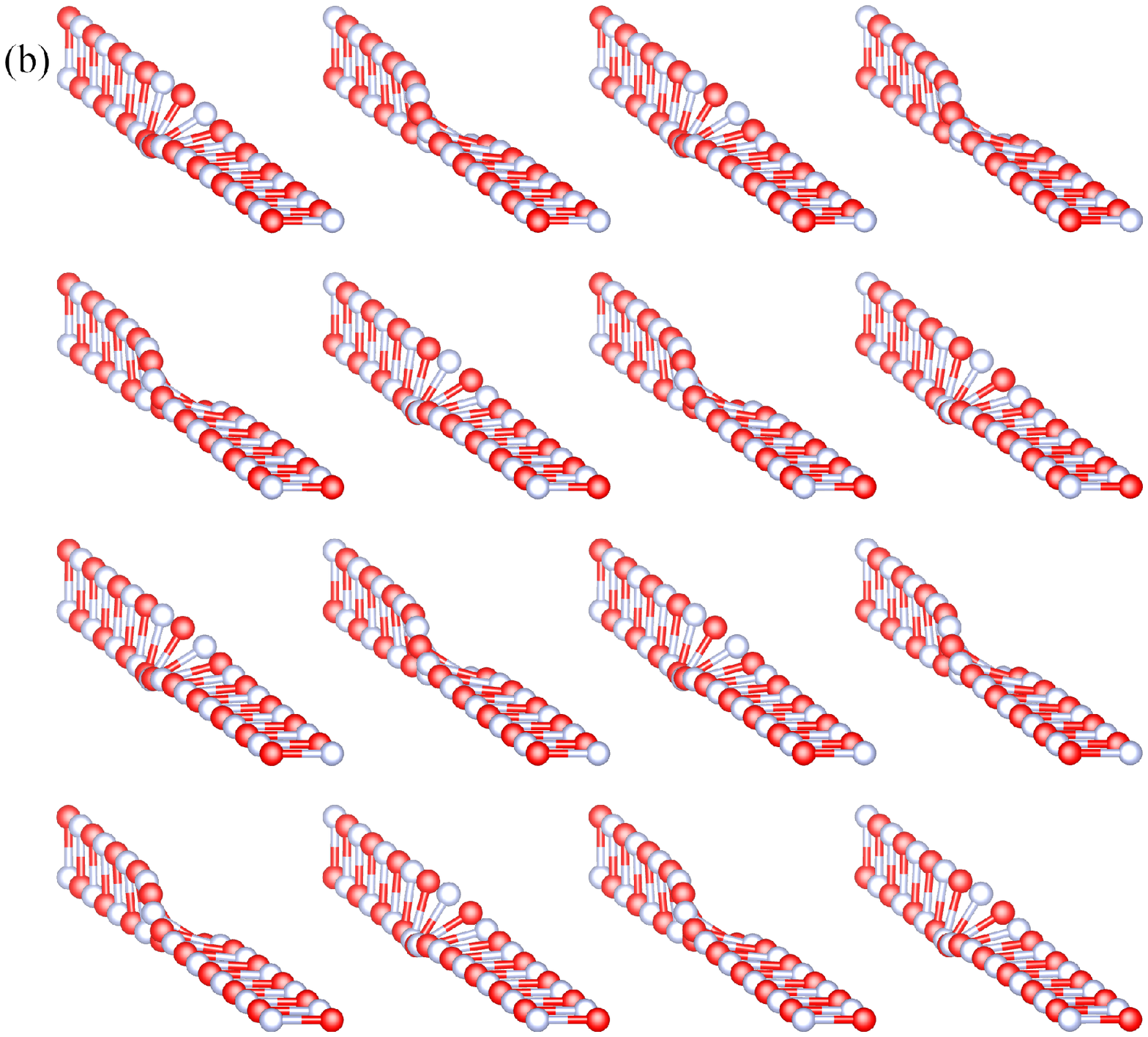}
\caption{Domain boundary profile when the dipoles have a finite spatial size. Here we perform simulations similar to those used in Fig.~\ref{Fig:BoundaryDipoleProfile}, but with 50 layers. Panel (a) shows the orientation angle as a function of layer index for the cases where the dipole size is given by $r_d/r_0=0.001$ (red squares), 0.01 (green circles) and 0.1 (blue triangles). The larger the dipole size, the sharper the boundary. Panel (b) shows the dipole configuration in a 20-layer sample with $r_d/r_0=0.1$, i.e.~the configuration that corresponds to the blue triangles in Panel (a).}
\label{Fig:BoundaryDipoleProfileForFiniteDipoleSize}
\end{figure}

As we have mentioned above, the infinite degeneracy of the optimal configuration is lifted if we consider dipoles of finite size, i.e.~a finite distance between the two charges of the dipole. In that case, the optimal configuration is six-fold degenerate: there is a preference for the dipoles to point along one of the crystal axes (x, y or z). This energetic preference also affects domain boundaries. In particular, one can expect a competition between the tendency favored by the dipole-dipole interaction term to have a gradual change in dipole orientations across the domain boundary and the tendency favored by the higher-order terms to align the dipoles with one of the three crystal axes. In Fig.~\ref{Fig:BoundaryDipoleProfileForFiniteDipoleSize} we show results for the dipole orientation angles in the case when the two domain orientations point in the y and z directions. Here we calculate the electrostatic forces and energies using Coulomb interactions between the individual charges of the dipoles, which is a more accurate approach for large dipoles. When the dipole size $r_d = 0.001 r_0$ and using a simulation with 50 layers (with 6 fixed layers and 44 unconstrained layers), we can see that the boundary is smooth, with barely any deviation from the linear increase of orientation angle as a function of position. When $r_d = 0.01 r_0$, we see a small deviation from the linear change in orientation angle, suggesting that a simulation with many more dipole layers would probably give a boundary width $\sim 10^2$. When $r_d = 0.1 r_0$, the boundary becomes quite sharp, with only a few layers where the dipole orientations deviate significantly from the orientations away from the boundary.

In order to provide experimental context to the results presented above, we give the numerical values for some of the relevant parameters in MAPbI$_3$. The lattice parameter for the cubic phase is $r_0=6.3$ \AA ~\cite{Baikie:2015aa}. The dipole moment of the MA molecule in the perovskite is $\sim 2$ D \cite{Madjet:2016ab}, and the N-C bond length is $\sim 1.46$ \AA ~\cite{Weller:2015ab}. Using this bond length as an estimate for $r_d$, we find that $r_d/r_0=0.23$, which is larger than the $r_d/r_0$ values used in Fig.~\ref{Fig:BoundaryDipoleProfileForFiniteDipoleSize}. As a result, the ratio $r_d/r_0$ in MAPbI$_3$ is sufficiently large to result in a sharp domain boundary. Sharp inter-domain interfaces have been suggested to enhance the transport of charge carriers by 3-4 orders of magnitude in comparison with the bulk conductivity\cite{Rashkeev:2015aa}. However, the contribution of the inorganic lattice may result in a different interfacial structure as a result of the competition between different interaction terms.

\section{Conclusion}
\label{Sec:Conclusion}

We have performed simulations of boundary regions between domains of different dipolar order in a simple cubic lattice of dipoles. Our results show that in the limit of point dipoles, it is energetically favorable to have wide boundaries with a smooth transition from the orientation of one domain to that of the other. In this case we find that as the boundary width is increased the energy of the boundary quickly decreases below the energy scale of single-dipole excitations. We also find that domains with dipole orientations perpendicular to a domain boundary result in higher boundary energy. When the spatial size of the dipole is a considerable fraction of the lattice parameter, it is energetically favorable to have sharp boundaries that extend only a few atomic layers.

Our result help provide insight and understanding of the expected properties of domain boundaries in MAPbI$_3$ and can form the basis for further studies on the dynamics and thermodynamics of dipolar domains in MAPbI$_3$. An accurate model of MAPbI$_3$ would include the inorganic lattice formed by the Pb and I atoms, and deviations from cubic order in this lattice can in fact play an important role in restricting the dipole orientations to a few energetically favorable directions. Our results then enter the problem as one of several physical mechanisms that together determine full details of the dipole configurations in the material.

We finally emphasize that our analysis and results can be applied to any dipole lattice governed by dipole-dipole interactions, including for example magnetic dipole lattices.

\appendix*
\section{Optimal configuration energy}
\label{Sec:Appendix}

\setcounter{figure}{0}
\renewcommand\thefigure{A\arabic{figure}}

Here we make two comments on the energy of the optimal configuration. First, to be self-contained we explain the origin of the infinite degeneracy of the Luttinger-Tisza optimal configuration \cite{Luttinger:1946aa}. As mentioned in the main text, in the optimal configuration all the dipoles are determined by the dipole vector at the origin, which is given by $(p_x, p_y, p_z)$: the dipole vector at location $(x, y, z)$ is given by $\left( [-1]^{y+z} p_x, [-1]^{x+z} p_y, [-1]^{x+y} p_z \right)$. Because of the cubic symmetry, for every dipole at location $(x, y, z)$ there are dipoles at locations given by the permutations [e.g.~$(z, y, x)$] as well as mirror image locations [e.g.~$(x,-y,z)$]. This property has the following implications for the two terms in the dipole-dipole interaction energy. First, taking the dipoles at the origin and at $(x, y, z)$, the first term in the energy has the dot product
\begin{equation}
\vec{p}_i\cdot\vec{p}_j = (-1)^{y+z} p_x^2 + (-1)^{x+z} p_y^2 + (-1)^{x+y} p_z^2.
\end{equation}
\\
The coefficient of each one of these three terms is either 1 or $-1$, depending on the coordinates $(x, y, z)$. For every such term, there are partners that are obtained from the permutations of the coordinates, such that we can regroup the terms in the sum over all the permutations into a sum over terms that are given by either $p_x^2 + p_y^2 + p_z^2$ or $-\left(p_x^2 + p_y^2 + p_z^2\right)$. Since $p_x^2 + p_y^2 + p_z^2$ is a constant, this sum will be independent of the orientations of the dipoles. For the second term in the energy, taking the dipoles at the origin and at $(x, y, z)$, we find the product
\begin{widetext}
\begin{eqnarray}
\left(\vec{p}_i\cdot\vec{r}_{ij}\right) \left(\vec{p}_j\cdot\vec{r}_{ij}\right) & = & \left( x p_x + y p_y + z p_z \right) \left( (-1)^{y+z} x p_x + (-1)^{x+z} y p_y + (-1)^{x+y} z p_z \right) \nonumber \\
& = & (-1)^{y+z} x^2 p_x^2 + (-1)^{x+z} y^2 p_y^2 + (-1)^{x+y} z^2 p_z^2 + \nonumber \\
& & (-1)^{x+z} x y p_x p_y + (-1)^{x+y} x z p_x p_z + (-1)^{y+z} x y p_x p_y + \nonumber \\
& & (-1)^{x+y} y z p_y p_z + (-1)^{y+z} x z p_x p_z + (-1)^{x+z} y z p_y p_z
\end{eqnarray}
\\
The cross terms, i.e.~the last six terms, cancel because for every nonzero coordinate, e.g.~$x$, there will be a partner term with the coordinate $-x$ giving the same term with the opposite sign. The first three terms in the sum obey an argument similar to the one that we gave above for $\vec{p}_i\cdot\vec{p}_j$: we can rearrange the terms in the sum into groups that have the form $\pm u^2 \left(p_x^2 + p_y^2 + p_z^2\right)$, where $u$ is one of the coordinates $(x, y, z)$ in the set of terms under consideration. Once again, since $p_x^2 + p_y^2 + p_z^2$ is a constant, this sum will be independent of the orientations of the dipoles. We therefore find that the total energy must be independent of the orientations of the dipoles and must hence be infinitely degenerate.

Secondly, one might wonder about the convergence of the total energy if we increase the cutoff distance for the dipole-dipole interaction energy. In an ordered system, the energies of all dipoles are equal. We therefore consider one dipole in the lattice. The energy for this single dipole is the sum of all the dipole-dipole interaction energies of this dipole with neighbouring dipoles up to the interaction cutoff distance:
\begin{equation}
E = \frac{E_0}{2} \sum_{\scriptsize\begin{array}{c} \vec{r}=(x,y,z) \\ x,y,z=0, 1, ... \\ 0 < \left|r\right|\leq r_c/r_0\end{array}} \frac{\vec{p}_{\vec{0}}\cdot\vec{p}_{\vec{r}} - 3 \left(\vec{p}_{\vec{0}}\cdot\hat{r}\right) \left(\vec{p}_{\vec{r}}\cdot\hat{r}\right)}{\left|r\right|^3}.
\label{Eq:SingleDipoleEnergy}
\end{equation}
For purposes of analyzing convergence properties, the sum of the energy contributions from all the neighbouring dipoles can be approximated by an integral. The integral runs over the renormalized distance variable $r$, ranging from zero (or rather a small value $\delta$ to avoid the divergence at zero) to the dimensionless cutoff distance $r_c/r_0$:
\begin{equation}
E \approx \frac{E_0}{2} \int_{\delta}^{r_c/r_0} dr \int_0^{\pi} d\theta \int_0^{2\pi} d\phi \left[ r^2 \frac{\vec{p}_{\vec{0}}\cdot\vec{p}_{\vec{r}} - 3 \left(\vec{p}_{\vec{0}}\cdot\hat{r}\right) \left(\vec{p}_{\vec{r}}\cdot\hat{r}\right)}{r^3} \right].
\label{Eq:SingleDipoleEnergyIntegral}
\end{equation}
\end{widetext}
At each point in the integral, the integrand is given by $1/r$ multiplied by the projections of the average dipole orientation at distance $r$ along $\vec{p}_{\vec{0}}$ and along $\hat{r}$. The integral $\int_{\delta}^{r_c/r_0}dr/r$ alone diverges logarithmically as $r_c/r_0\rightarrow\infty$. However, since the relevant configurations have a zero net polarization over any large volume, fluctuations will decrease as the square root of the volume under consideration, i.e.
\begin{equation}
dr d\theta d\phi \left| \vec{p}_{\vec{0}}\cdot\vec{p}_{\vec{r}} - 3 \left(\vec{p}_{\vec{0}}\cdot\hat{r}\right) \left(\vec{p}_{\vec{r}}\cdot\hat{r}\right) \right| < {\rm const.} \times r^{-1/2} dr d\theta d\phi.
\label{Eq:SingleDipoleEnergyIntegral}
\end{equation}
The integrand in Eq.~(\ref{Eq:SingleDipoleEnergyIntegral}) therefore approaches zero faster than $1/r$, and the integral converges. Indeed, numerical calculations show that the energy per dipole in the optimal configuration (in units of $E_0$) oscillates with increasing cutoff distance and remains between $-3$ and $-2.5$ for all values of $r_c/r_0$ above 2.5, as shown in Fig.~\ref{Fig:EnergyAsFunctionOfInteractionRange}. For the special case $r_c/r_0=3$, which we use in several calculations presented in the main text, by numerically performing the sum in Eq.~(\ref{Eq:SingleDipoleEnergy}) with $r_c=3r_0$, we find that the total energy of the central dipole is $-2.79E_0$.


This work was made possible by NPRP grant \# 8-086-1-017 from the Qatar National Research Fund (a member of Qatar Foundation). The findings achieved herein are solely the responsibility of the authors.


%

\end{document}